\documentclass[12pt,preprint,a4paper]{aastex}
\usepackage{times}

\usepackage{graphics,epsf}
\usepackage{amsmath}                
\usepackage{amsfonts}               
\usepackage{amssymb}                
\usepackage{epsfig}                 
\usepackage{graphicx}
\usepackage{subcaption}

\def \cm{~\rm{cm}}
\def \s{~\rm{s}}
\def \km{~\rm{km}}

\def \K{~\rm{K}}
\def \g{~\rm{g}}

\begin{document}

\title{A PLANAR JITTERING-JETS PATTERN IN CORE COLLAPSE SUPERNOVA EXPLOSIONS}

\author{Oded Papish\altaffilmark{1} and Noam Soker\altaffilmark{1}}

\altaffiltext{1}{Department of Physics, Technion -- Israel Institute of Technology, Haifa
32000, Israel; papish@techunix.technion.ac.il; soker@physics.technion.ac.il}

\begin{abstract}
We use 3D hydrodynamical numerical simulations and show that jittering bipolar jets that power core-collapse supernova (CCSN) explosions
channel further accretion onto the newly born neutron star (NS) such that consecutive bipolar jets tend to be launched in
the same plane as the first two bipolar jet episodes.
In the jittering-jets model the explosion of CCSNe is powered by jittering jets launched by an intermittent accretion disk
formed by accreted gas having a stochastic angular momentum.
The first two bipolar jets episodes eject mass mainly from the plane defined by the two bipolar axes. Accretion then proceeds from
the two opposite directions normal to that plane. Such a flow has an angular momentum in the direction of the same plane.
If the gas forms an accretion disk,
the jets will be launched in more or less the same plane as the one defined by the jets of the first two launching episodes.
The outflow from the core of the star might have a higher mass flux in the plane define by the jets.
In giant stellar progenitors we don't expect this planar morphology to survive as the massive hydrogen envelope will tend to
make the explosion more spherical. In SNe types Ib and Ic, where there is no massive envelope, the planar morphology might
have an imprint on the supernova remnant.
We speculate that planar jittering-jets are behind the morphology of the Cassiopeia A supernova remnant.
\end{abstract}

\section{INTRODUCTION}
\label{sec:intro}

One class of core collapse supernova (CCSN) explosion models is based on neutrino \citep{Colgate1966}, mainly the delayed neutrino mechanism (e.g.,
\citealt{bethe1985,Burrows1985,Burrows1995,Fryer2002,Ott2008,Marek2009,Nordhaus2010,Kuroda2012,Hanke2012,Janka2012,Bruenn2013}).
However, recent 3D numerical studies have shown that the desired explosions are harder to achieve \citep{Couch2013, Jankaetal2013, Takiwakietal2013}
than what 2D numerical simulations had suggested (for a summary of problems of the delayed neutrino mechanism see \citealt{Papishetal2014}).
The problems of the delayed-neutrino mechanism can be overcome if there is a strong wind, either from an accretion disk \citep{kohri2005}
or from the newly born neutron star (NS). Such a wind is not part of the delayed-neutrino mechanism, and most researchers consider this wind
to have a limited contribution to the explosion.

Another class of explosion mechanisms is the jittering-jet scenario \citep{Soker2010, Papish2011, Papish2012a, Papish2012b, PapishSoker2014, GilkisSoker2013}.
Processes for CCSN explosion by jets were considered before the development of the jittering-jet scenario
(e.g. \citealt{LeBlanc1970, Meier1976, Bisnovatyi1976, Khokhlov1999, MacFadyen2001,Hoflich2001, Woosley2005, Burrows2007, Couch2009,Couch2011,Lazzati2011}).
However, most of these MHD models require a rapidly spinning core before collapse starts, and hence are limited to a small fraction of all CCSNe.
The jittering-jet scenario posits that {\it all CCSNe are exploded by jets}.
Recent observations (e.g. \citealt{Milisavljevic2013,Lopez2013,Ellerbroek2013}) show indeed that jets might have a much more
general role in CCSNe than what is expected in the neutrino-driven mechanisms and mechanisms that require rapidly rotating cores.

In the jittering-jets scenario the sources of the angular momentum for disk formation are the convective regions in the core \citep{GilkisSoker2013}
and instabilities in the shocked region of the collapsing core, e.g., neutrino-driven convection or the standing accretion shock instability (SASI).
Recent 3D numerical simulations show indeed that neutrino-driven convection and SASI are well developed in the first second after core bounce
\citep{Hankeetal2013, Takiwakietal2013} and the unstable spiral modes of the SASI can amplify magnetic fields \citep{Endeveetal2012}.
The spiral modes with the amplification of magnetic fields build the ingredients necessary for jets' launching.

When the average specific angular momentum of the matter in the pre-collapse core is small relative to the amplitude of the specific angular momentum
of these instabilities, intermittent jets-launching episodes with random directions occur.
The two launching axes of the first two launching episodes define a plane.
Using the FLASH numerical code we set a numerical study of the accretion pattern that is likely to be formed after the first two launching episodes.
The code and numerical set-up for the 3D simulations are described in section \ref{sec:setup}.
The accretion pattern following two jets-launching episodes, followed by a third episode, is described in section \ref{sec:accretion}.
Our summery is in section \ref{sec:summary}.

\section{NUMERICAL SETUP}
\label{sec:setup}

We study the accretion structure that result from multiple jets-launching episodes using the  {\sc flash} gasdynamical numerical code version 4.2 \citep{Fryxell2000}.
The widely used {\sc flash} code is a publicly available code for supersonic flow suitable for astrophysical applications.

\begin{figure}[h]
        \centering
                \centering
                \includegraphics[width=0.8\textwidth]{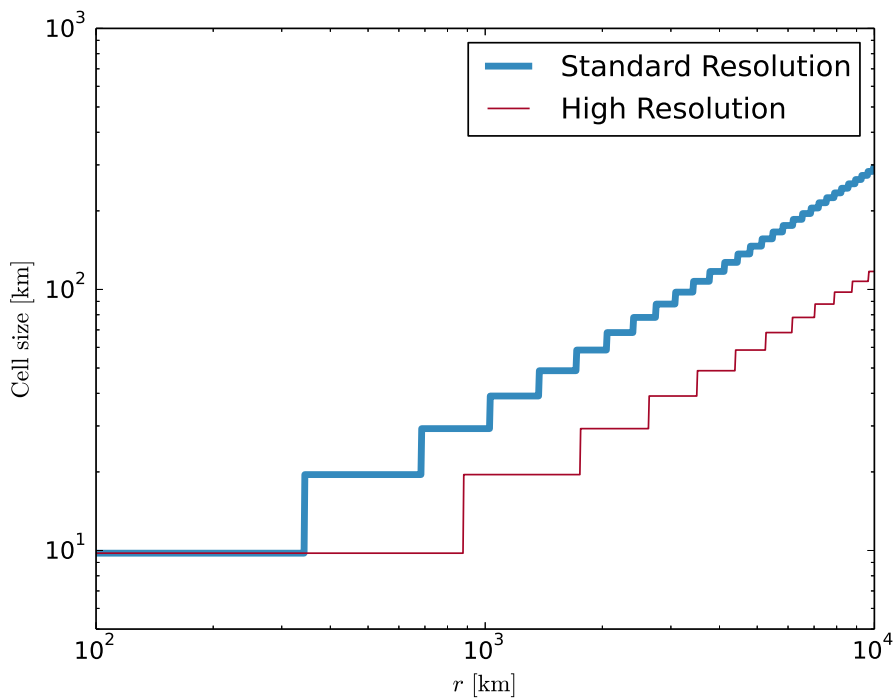}

\caption{ {{{  {The dependence of the grid cell size on the distance from the center of the simulation grid. The thin line represent the simulation used to check the convergence of our results.} }}} }
\label{fig:res}
\end{figure}

The simulations are done using the split PPM solver of {\sc flash}. We use 3D Cartesian coordinates with an adaptive mesh refinement (AMR) grid.
{{{{ Fig. \ref{fig:res} shows the resolution in our simulations as a function of distance from the center. In the inner part the cells size is 10 km, increasing with radius.} }}}

{{{{We treat the spherical inner region of up to $100 \km$ from the center as a hole.} }}} This means that we are not simulating the NS itself,
nor the assumed accretion disk. The boundary condition at the edge of the hole is inflow only,
meaning the velocity cannot be positive in the radial direction unless we inject a jet at that specific zone.

{{{ { To check the sensitivity of our simulations to the resolution used we run a test case with higher resolution (see Fig. \ref{fig:res}). A compression with the higher resolution run, that will be presented in section \ref{subsec:res} (Figs. \ref{fig:acc}, \ref{fig:high}), shows that the low and high resolutions runs give results that are within $\sim 10 \%$ from each other.  } }}}

We start the simulations after the core has collapsed and bounced back. The initial conditions are taken for a $15 M_\odot$ model
from the 1D simulations of \cite{Liebend2005} at a time of $t \simeq 0.2 \s$ after bounce.
We map their results into 3D, including the chemical composition. {{{{ We did not included nuclear reactions in our calcualtions. } }}}
We set outflow boundary conditions at the exteriors of the simulation's domain.

In each episode we inject one pair of two opposite jets (bipolar jets).
All jets' axes are in one plane, that we take to be the $y=0$ plane of the numerical Cartesian grid.
We take the $z$ axis of the numerical grid  to be at $40^\circ$ to the direction of the symmetry axis of the first jets pair.
The $n$ launching episode results in two opposite jets at some angle $\theta_n$ from the direction of the first jets.
All jets have an initial conical shape, with a half opening angle of $10^\circ$.

\section{ACCRETION PATTERN}
\label{sec:accretion}
\subsection{Simulated Cases}
\label{cases}
\begin{figure}[h]
        \centering
                \centering
                \includegraphics[width=0.8\textwidth]{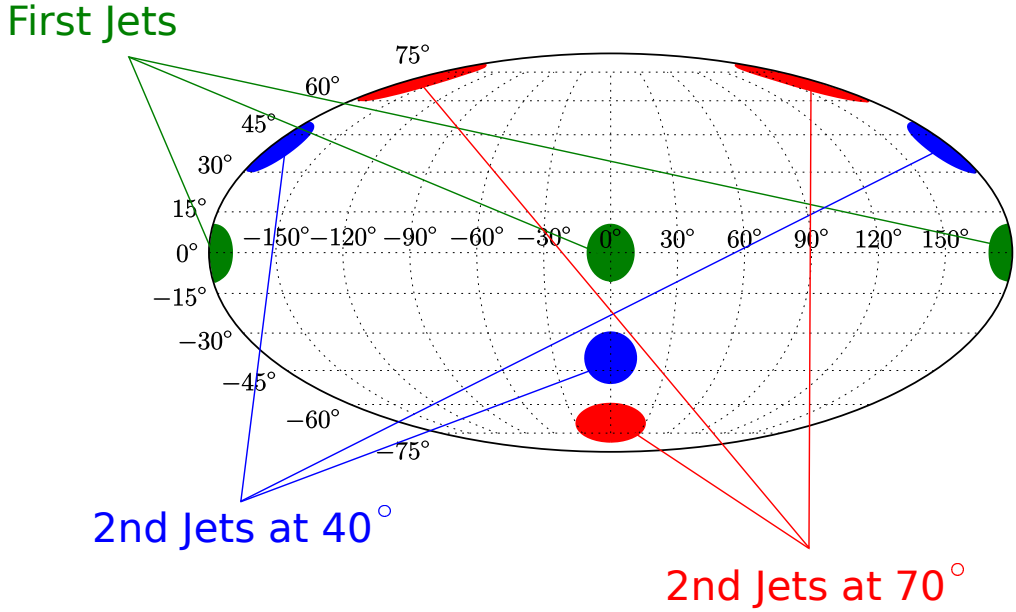}

\caption{The simulations grid in the Mollweide projection. Shown are the directions of the first and second episodes in the different Runs.
In each episode two opposite jets are launched. One of the jets in each episode is represented by two halves in this diagram.}
\label{fig:grid}
\end{figure}

We start each simulation by injecting two opposite jets, the first jets' launching episode.
The second bipolar jets pair is injected either at an angle of $40^\circ$ or $70^\circ$ relative to the first episode.
Each jet launching episode lasts $0.05 \s$, and the second episode is launched immediately after the end of the first episode.
The different simulated cases are summarized in Table \ref{table}. The direction of the jets launched in the first two episodes are presented in Fig. \ref{fig:grid}. Note that in that presentation one of the jets in each pair is represented by two regions.
\begin{table}[h!]

    \begin{tabular}{llll} \hline \hline
    Run & $\theta_1$ & $\theta_2$ & $\theta_3$ \\
    Active($s$) & $0-0.05$ & $0.05-0.1$ & $0.1-0.15$ \\ \hline
    A1    & $0^\circ$ & $40^\circ$ & $80^\circ$          \\
    A2    & $0^\circ$ & $40^\circ$ & $-40^\circ$  \\
    B1    & $0^\circ$ & $70^\circ$ & $55^\circ$ \\
    B2    & $0^\circ$ & $70^\circ$ & $110^\circ$ \\\hline
    \end{tabular}
    \caption{A1, A2, B1, and B2 are the three different simulated cases.
    The angles $\theta_{2}$ and
    $\theta_{3}$ are the angles of the jets in the second and third jets' launching
episodes relative to the direction of the first jets that are injected at $40^\circ$ from
    the $z$ axis. All jets are injected in the $y=0$ plane.
Each episode lasts $0.05 \s$.}
    \label{table}
\end{table}

\subsection{First jet-launching episode}
\label{first}
\begin{figure}[h!]
        \centering
                \includegraphics[width=\textwidth]{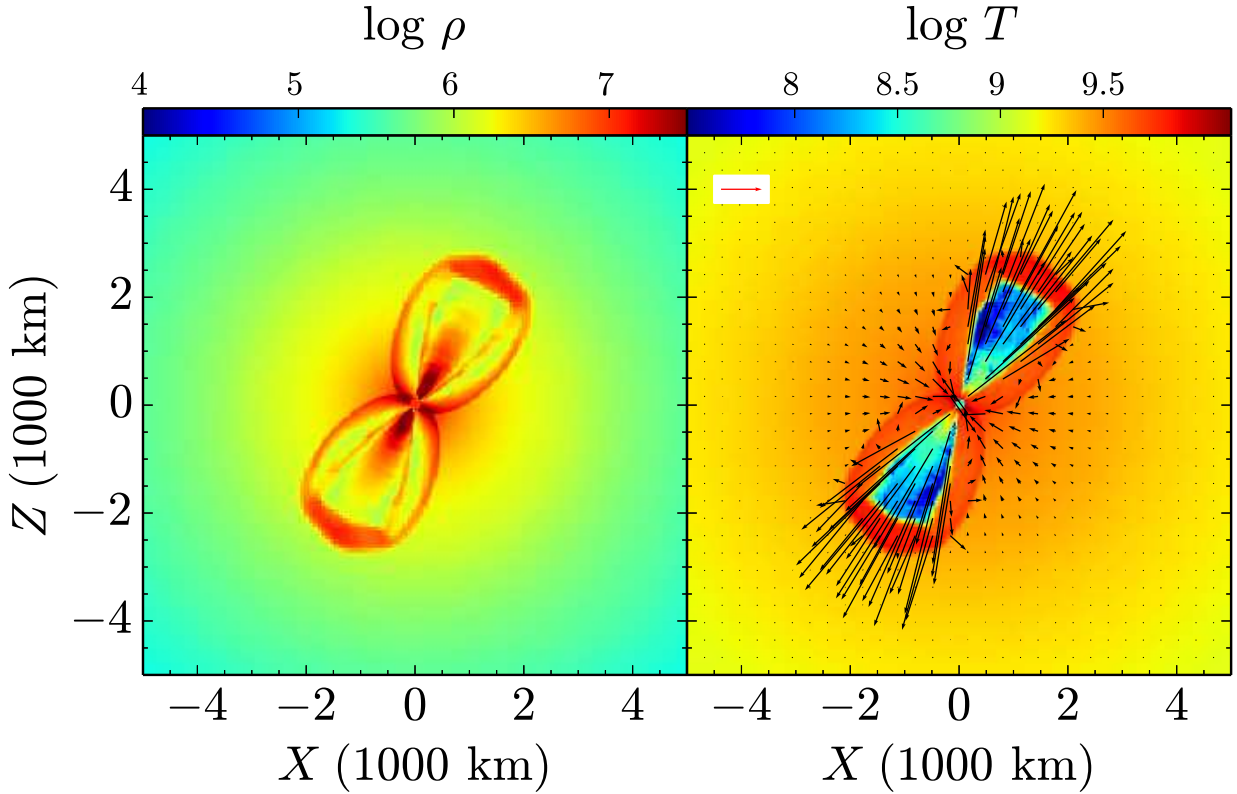}
         \caption{Flow pattern in the $y=0$ plane at the end of the first jets launching episode ($t=0.05 \s$), common to all simulated cases.
         Left Panel: density, with a color coding in logarithmic scale and units of $\g \cm^{-3}$.
         Right Panel: temperature in log scale and in units of $\K$, and a velocity map. Velocity is proportional to
         the arrow length, with inset showing an arrow for $30,000 \km \s^{-1}$. }
        \label{fig:dens-temp-05}
\end{figure}

In Fig. \ref{fig:dens-temp-05} we present the density, left panel, and the velocity and temperature, right panel, maps at the end
of the first jets launching episode.
The flow structure has the typical structure of jet-inflated bubbles (see review by \citealt{Sokeretal2013}). Here we should note the following.
($i$) The jets shock material to high temperatures and densities, where nucleosynthesis takes place.
These regions have high velocities and can have imprint on the distribution of different isotopes in the SN remnant at later times.
This will not be studied here.
($ii$) The pre-jets flow is that of a collapsing core onto a newly formed NS.  After the first jets launching episode accretion
continues from directions near the plane perpendicular to the jets' axis.

\begin{figure}[h!]
        \centering
         \includegraphics{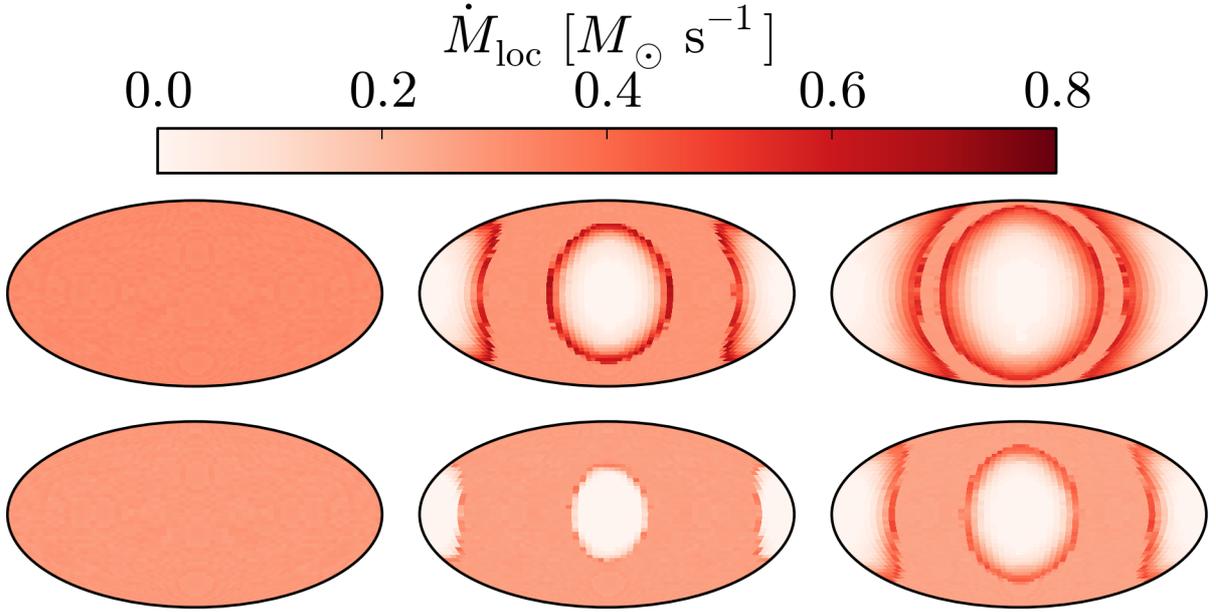}
\vspace*{1.0 cm}
\caption{The inflow mass flux at times $t=0, 0.026,0.05 \s$, i.e., at the beginning, middle and end of the first jets-launching episode,
given on spheres of radius $r= 500 \km$ (upper row) and $r= 1000 \km$ (lower row).
Mass inflow rate at each point is calculated as if the entire sphere would have the same inflow mass flux
as in the given point (eq. \ref{eq:phil}). White areas are regions with outflow, i.e., the jets.}
\label{fig:one_jets500}
\end{figure}

Let us elaborate on the accretion pattern.
In Fig. \ref{fig:one_jets500} we present the inflow mass flux on spheres of radii $500$ and $1000 \km$.
We present the local mass inflow rate $\dot M_{\rm loc}$, defined as if the entire sphere would have the same inflow mass flux as in the given location,
\begin{equation}
\dot M_{\rm loc}= 4 \pi r^2 \rho v_{\rm in},
\label{eq:phil}
\end{equation}
where $\rho$ and $v_{\rm in}$ is the density and the radial inward velocity at the point.
The first jets pair is able to penetrate thorough the inflowing matter to beyond $1000 \km$ \citep{PapishSoker2014}.
As the jets gas is in outflow, it appears as white areas in Fig. \ref{fig:one_jets500} and the following similar figures.

\subsection{A bipolar accretion pattern}
\label{bipolar}

\begin{figure}[h!]
        \centering
                \includegraphics[width=\textwidth]{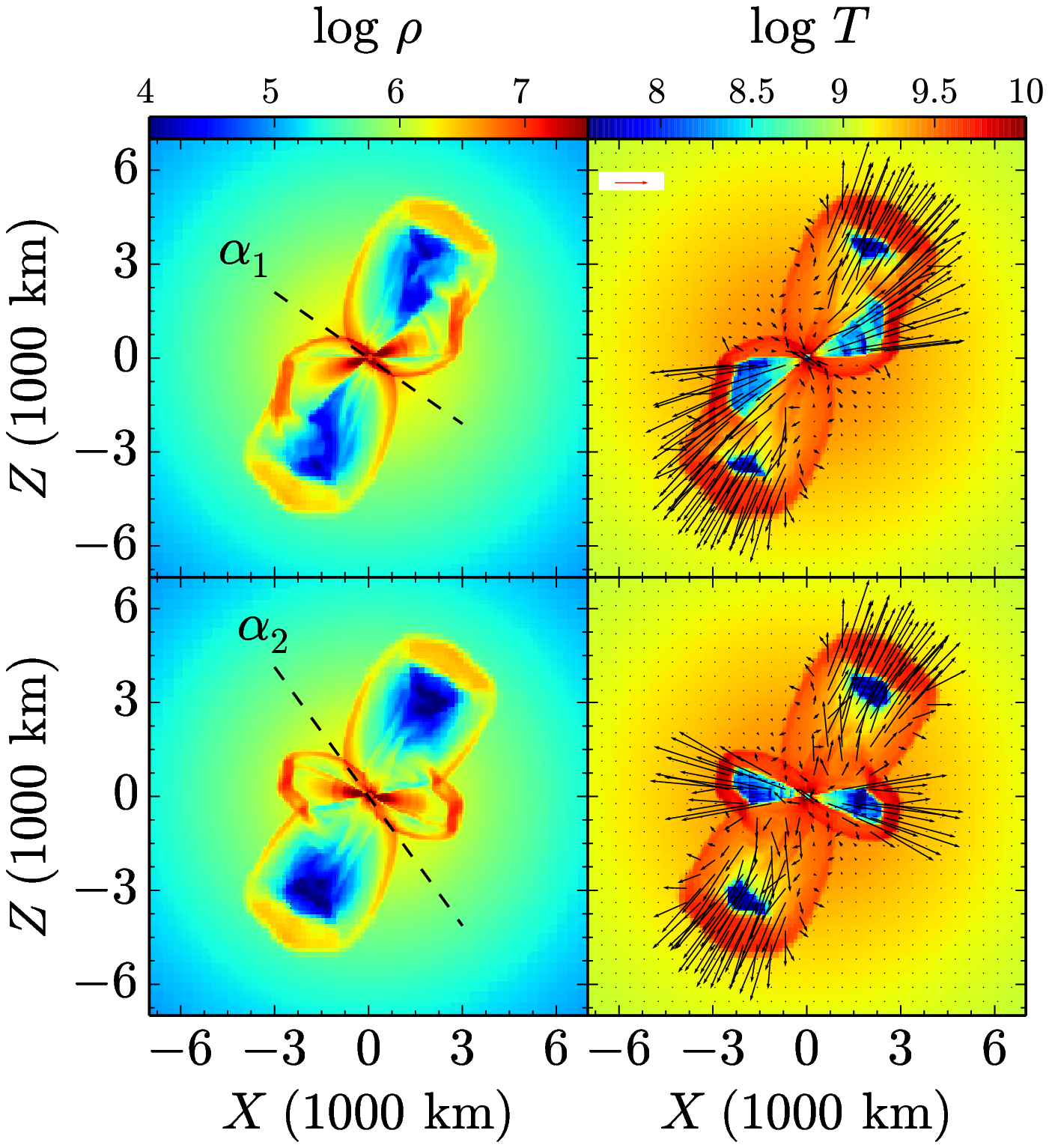}
                \caption{Flow structure at the end of the second jets episode at $t=0.1\s$ in the $y=0$ plane,
                for Run A in the upper panels, and for Run B in the lower panels.
                Left Panels: density, with a color coding in log scale and units of $\g \cm^{-3}$. The dashed line indicate the intersect of
                the plane shown in Fig. \ref{fig:global2s}.
                    Right Panels: temperature in log scale in units of $\K$, and velocity arrows.
                Velocity is proportional to the arrow length, with inset showing an arrow for $30,000 \km \s^{-1}$. }
        \label{fig:global2s}
\end{figure}

We run two cases of a second jets-launching episode, in directions of $40^\circ$ (Run A) and $70^\circ$ (Run B)
relative to the direction of the jets' axis in the first episode. All jets' axes are in the $y=0$ plane.
The density maps, left panels, and the temperature maps with velocity arrows, right panels, of the two cases
at the end of the second episode, $t=0.1 \s$, are presented in Fig. \ref{fig:global2s}.
All plots are in the $y=0$ plane. From these panels, two for each Run, we learn that the strongest inflow is from
two opposite directions, through which we draw a dashed line on the density maps of the figure at angles of $\alpha_1 = 35^\circ$, and $\alpha_2 = 54^\circ$ respectively relative to the x-axis.

\begin{figure}[h!]
\centering
\includegraphics[width=\textwidth]{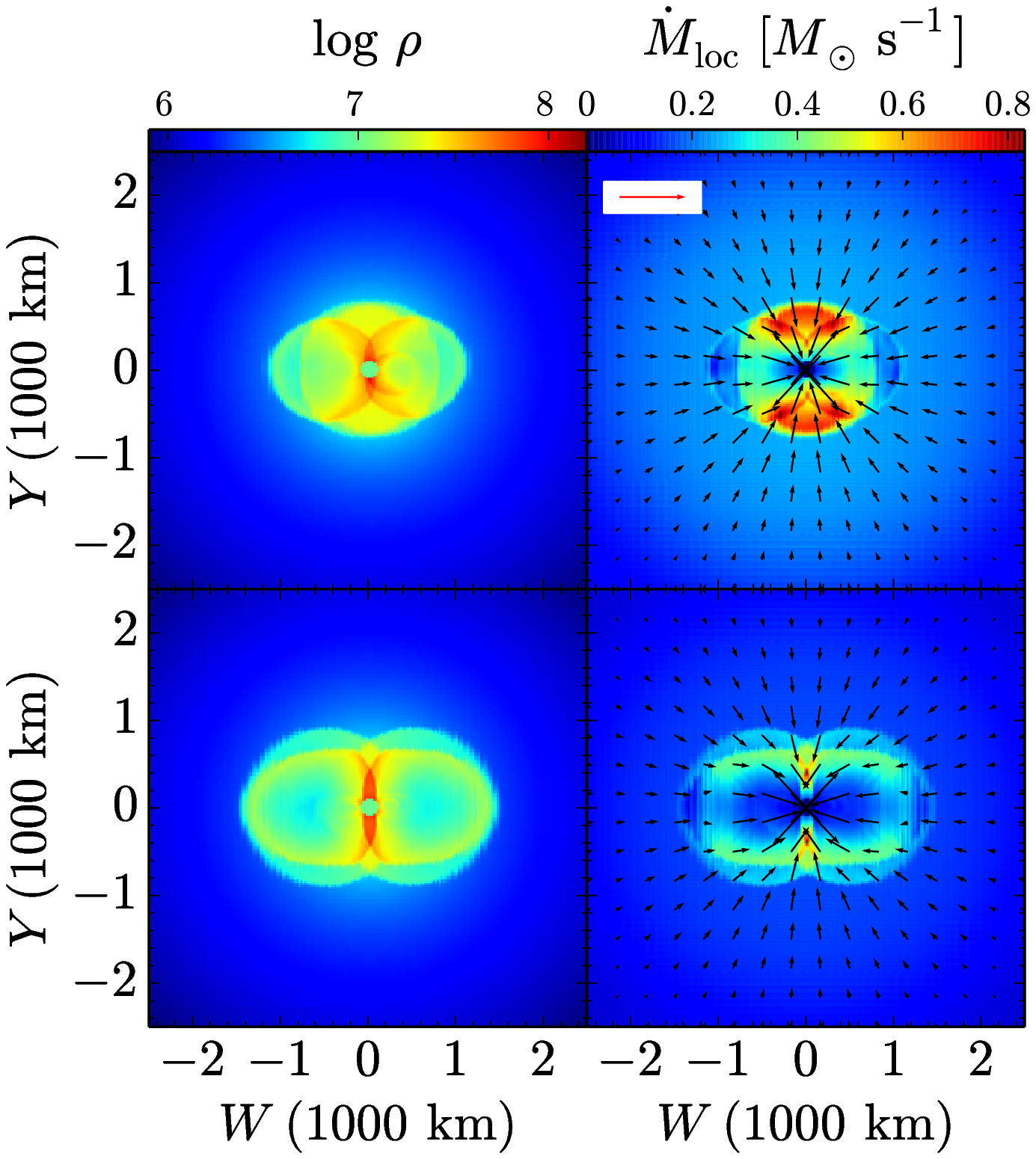}
\caption{Flow structure at the end of the second jets episode at $t=0.1\s$, as in Fig. \ref{fig:global2s}, but in a plane perpendicular to the $y=0$ plane and intersecting it at the dashed line drawn on the left panels of Fig. \ref{fig:global2s}. The coordinate along the dashed line is define as $W$, and it is along the line $z=-0.7x$ for Run A (upper panels) and $z=-1.38x$ for Run B (lower panels). Note that the panels here show a small inner region of the grid. Partition of panels as in Fig. \ref{fig:global2s}, but in the right panels the color represent mass influx rate $\dot M_{loc}$.}
        \label{fig:NormalPlane}
\end{figure}

To identify the inflow pattern of the gas, most of which will be eventually accreted by the NS, we present in Fig. \ref{fig:NormalPlane}
the flow structure in a plane perpendicular to the $y=0$ plane, and cutting it along the dashed lines
drawn in the density maps of Fig. \ref{fig:global2s}. Note that the two planes for Run A and Run B are not identical.
The left panels of Fig. \ref{fig:NormalPlane} present the density maps of the two Runs, while the right panels present the inflow mass flux
and flow velocity maps.
The mass flux is  $\dot M_{\rm loc}$ as define in equation (\ref{eq:phil}).
In is evident that the inflow close to the NS at the center is mainly from two general opposite directions, the $+y$ and $-y$ directions,
that are perpendicular to the axes of the two jets' launching episodes.
When the angle between the two first jet-pairs is small, $40^\circ$, the high inflow mass flux is from two extended opposite areas.
These becomes smaller when the angle between the two episodes are large, $70^\circ$.
A bipolar accretion flow pattern has emerged.

\begin{figure}[h!]
        \centering
             \includegraphics[width=\textwidth]{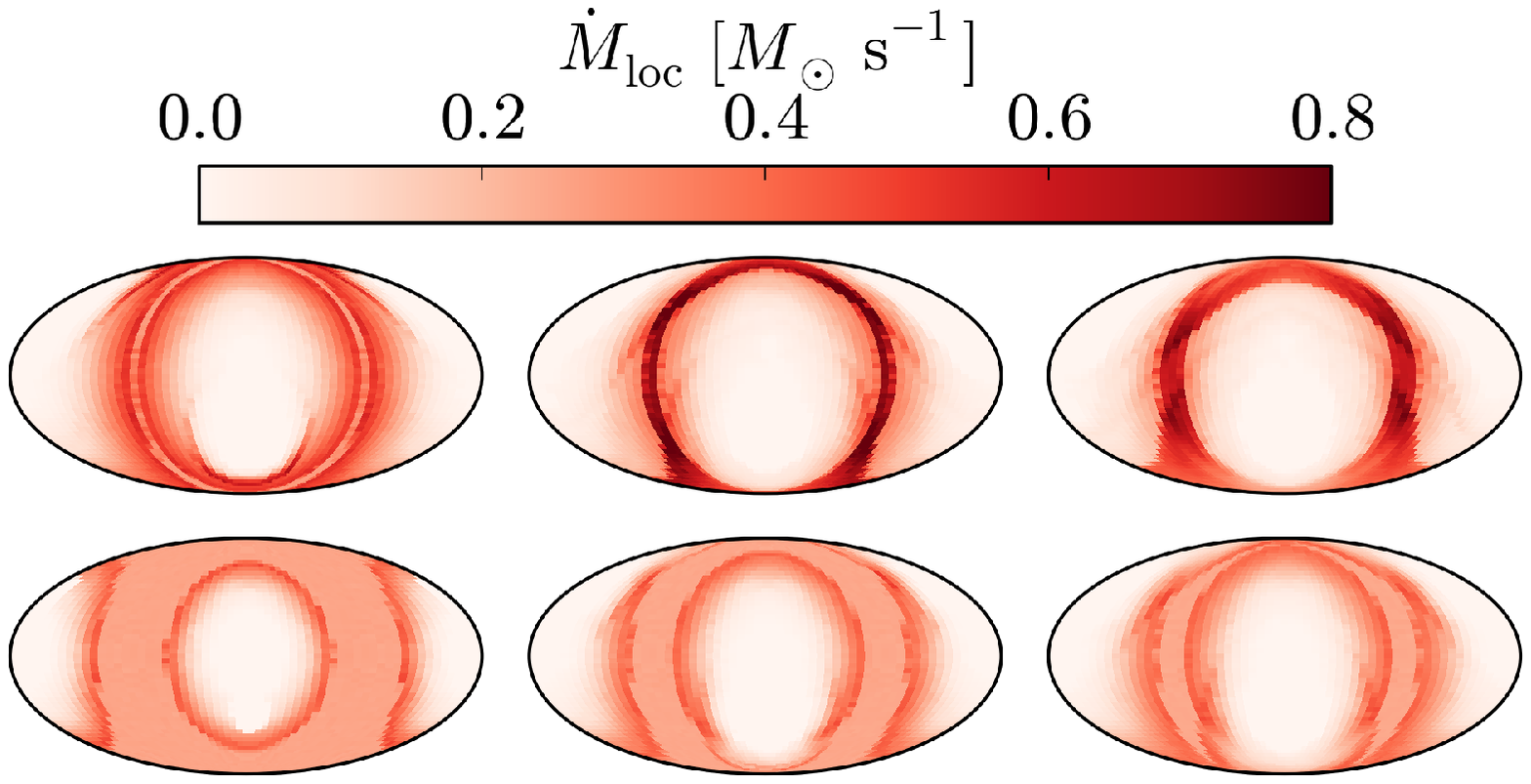}

\vspace*{1.0 cm}
\caption{The inflow mass flux at $t=0.062, 0.084, 0.1 \s$, from left to right, i.e., at the beginning, middle and end of the second
jets-launching episode, for Run A, given on spheres of radius $r= 500 \km$ (upper row) and $r= 1000 \km$ (lower row).
Mass inflow rate at each point is calculated as if the entire sphere would have the same inflow mass flux
as in the given point (eq. \ref{eq:phil}. White area are regions no inflow, some with outflow, i.e., the jets.
Note the emergence of two opposite high accretion rate regions. }
\label{fig:second_jets-40}
\end{figure}

\begin{figure}[h!]
        \centering
                   \includegraphics[width=\textwidth]{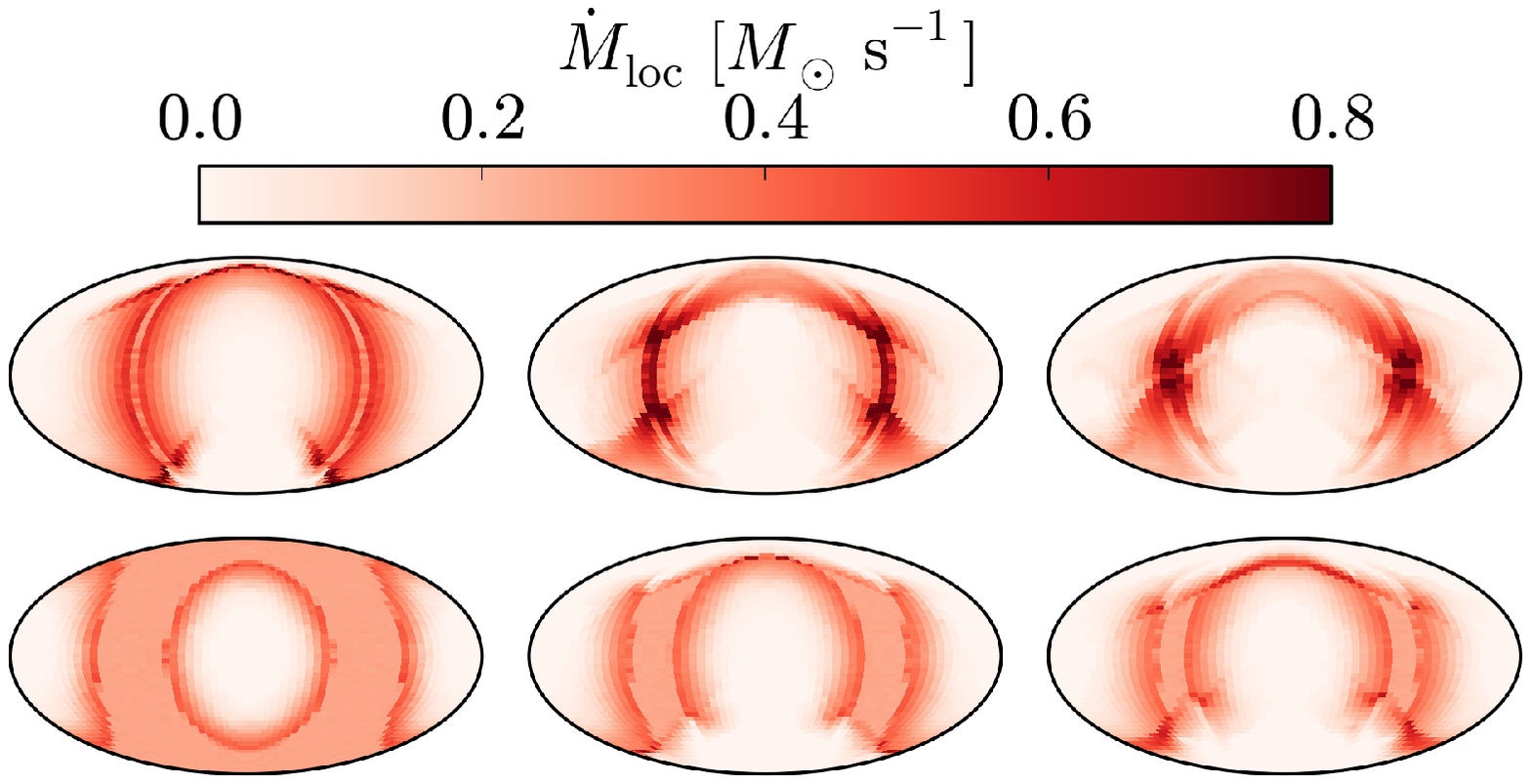}

\vspace*{1.0 cm}
\caption{Like Fig. \ref{fig:second_jets-40}, but for Run B, where the jets of the second episode are launched at
$\theta_1=70^\circ$ to the first jets' axis, rather than $\theta_1=40^\circ$. The bipolar accretion patter is clearly seen in the dark-red areas. }
\label{fig:second_jets-70}
\end{figure}

We can also use the Mollweide-projection (see Fig. \ref{fig:grid}) to present the emergence of the bipolar inflow (accretion) pattern.
This is shown in Fig. \ref{fig:second_jets-40} for Run A, and in Fig. \ref{fig:second_jets-70} for Run B.
Presented are the inflow mass fluxes on spheres of radii $500 \km$ and $1000 \km$, as in Fig. \ref{fig:one_jets500} for the first jets.
Again, the emergence of a bipolar inflow (accretion) structure is evident, particularly in Run B presented in Fig. \ref{fig:second_jets-70}.

\subsection{Later evolution}
\label{third}

\begin{figure}[h!]
        \centering
                \includegraphics[width=\textwidth]{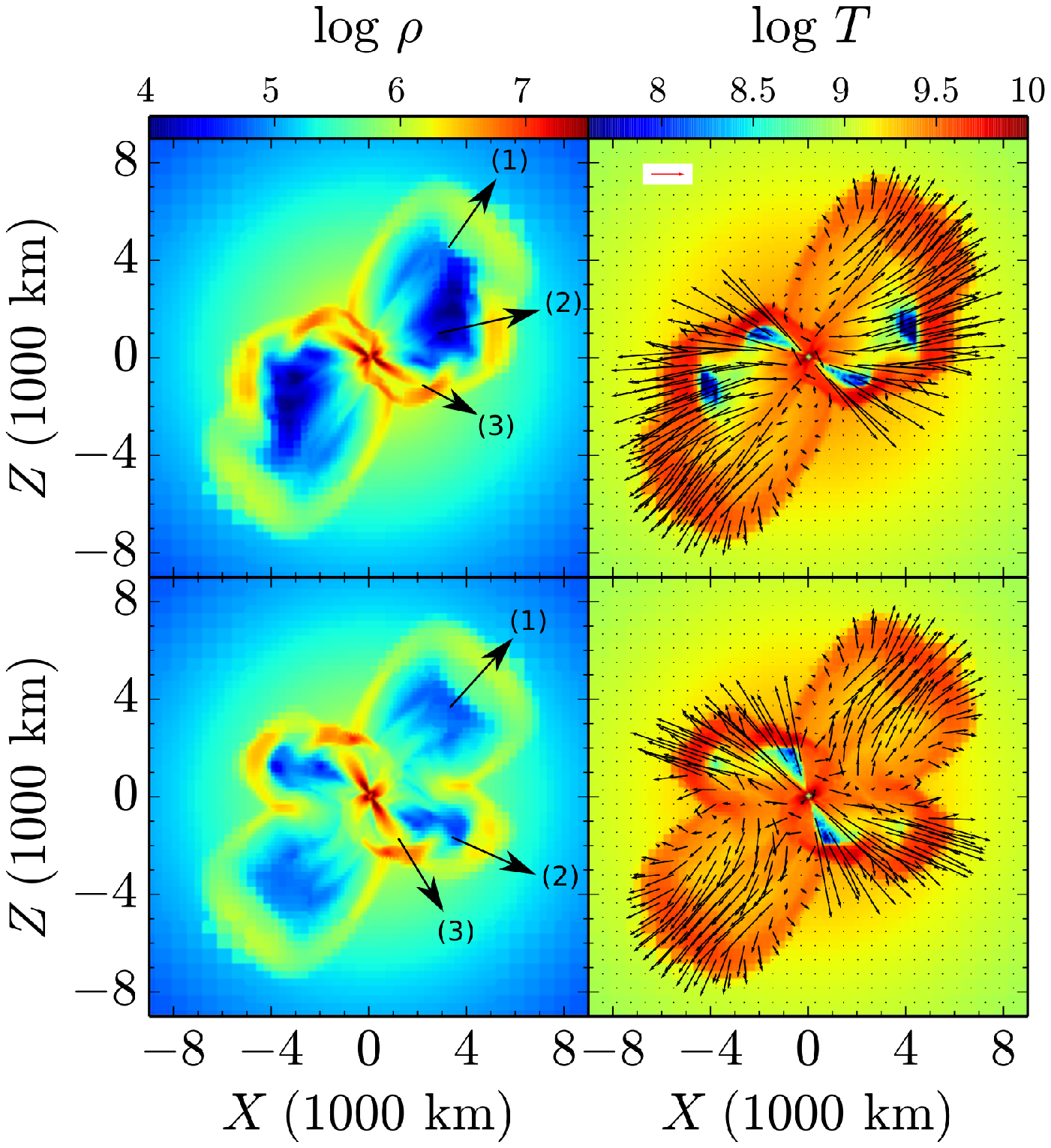}
                \caption{Flow structure in the $y=0$ plane at the end of the third jets launching episode, $t=0.15 \s$,
                for Run A1 (upper panels) and B2 (lower panels).
                Left Panels: density, with a color coding in logarithmic scale and units of $\g \cm^{-3}$. The three arrows depict the
                direction of jets' launching in the three episodes as numbered.
         Right Panels: temperature in log scale in units of $\K$, and velocity map. Velocity is proportional to
         the arrow length, with inset showing an arrow for $30,000 \km \s^{-1}$.
}
        \label{fig:third1}
\end{figure}

\begin{figure}[h!]
        \centering
                \includegraphics[width=\textwidth]{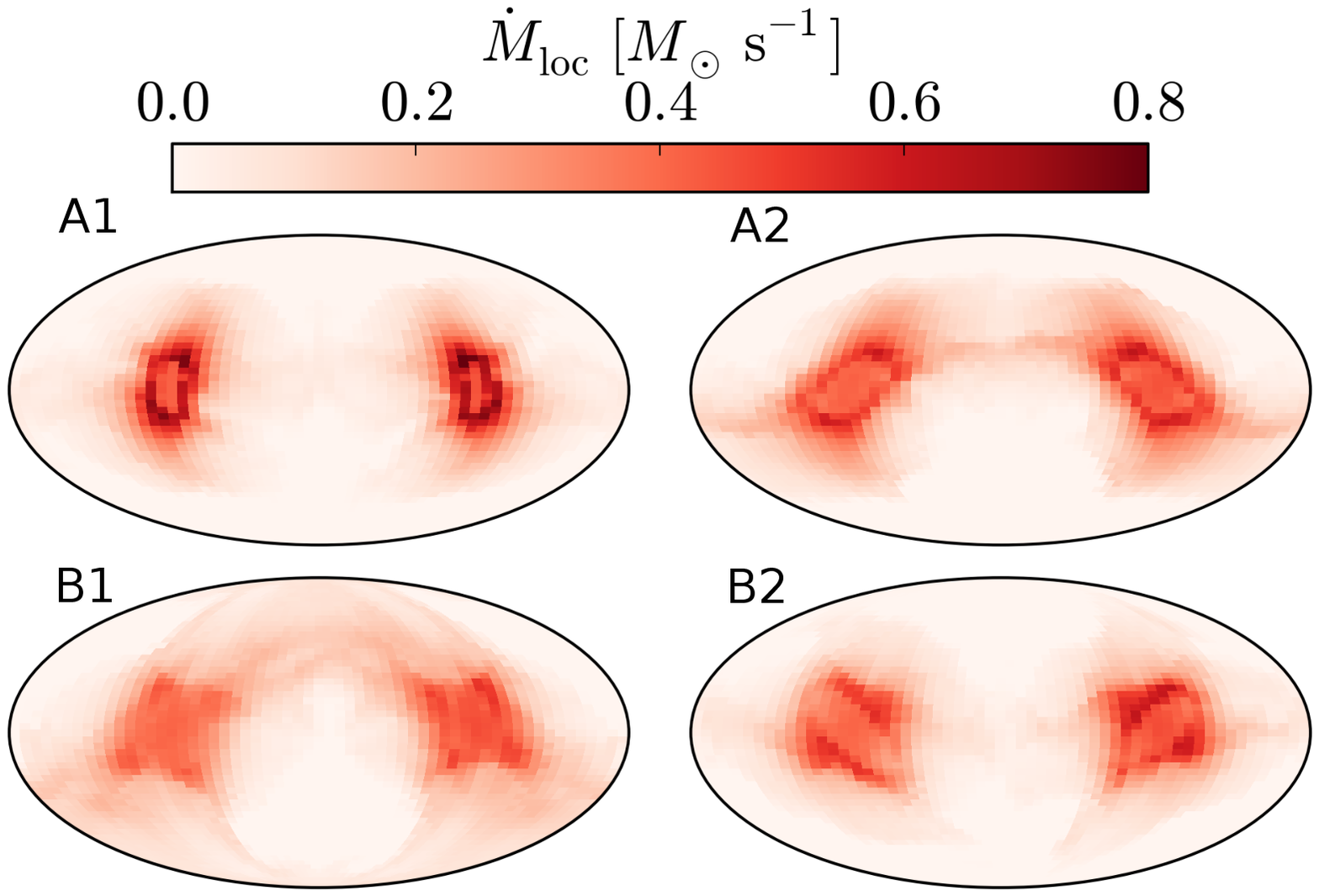}
                \caption{Like Fig. \ref{fig:second_jets-40} but at $t=0.15 \s$ and on a sphere of radius $r=500 \km$.
                The four cases are as indicated on the panels and according to Table \ref{table}.
                The third jets episode strengthens the bipolar accretion pattern.
                }
        \label{fig:NormalPlane-3}
\end{figure}

As we discussed in the next section, jets launched by accretion disks formed from the bipolar accretion pattern are
likely to be lunched perpendicular to the direction of accretion. Namely, they will be launched in, or close to, the $y=0$
plane defined by the jets' axes of the first two episodes.
We are not in the numerical stage to follow the angular momentum of the accreted gas and from that to
find the direction to launch the next jets episode.
We therefore arbitrarily run 4 cases for the third jets launching episode as listed in Table \ref{table}.
In all these cases the third jets launching episode is active in the time period $t=0.1-0.15 \s$.
We present the flow structure for two cases, A1 and B2, in Fig. \ref{fig:third1}, at the end of the
active phase $t=0.15 \s$.
In all cases we find that a third jets-launching episode strengthens the bipolar accretion pattern and the inflow becomes further
concentrated to two opposite directions.
In Fig. \ref{fig:NormalPlane-3} we use the Mollweide-projection (see Fig. \ref{fig:grid}) to present the inflow rates through a spheres of radius $r=500 \km$ for the four Runs at the end of the third episode. The jets of the third episode make the bipolar accretion pattern more prominent.

\begin{figure}[h!]
\begin{center}
\includegraphics[width=1\textwidth]{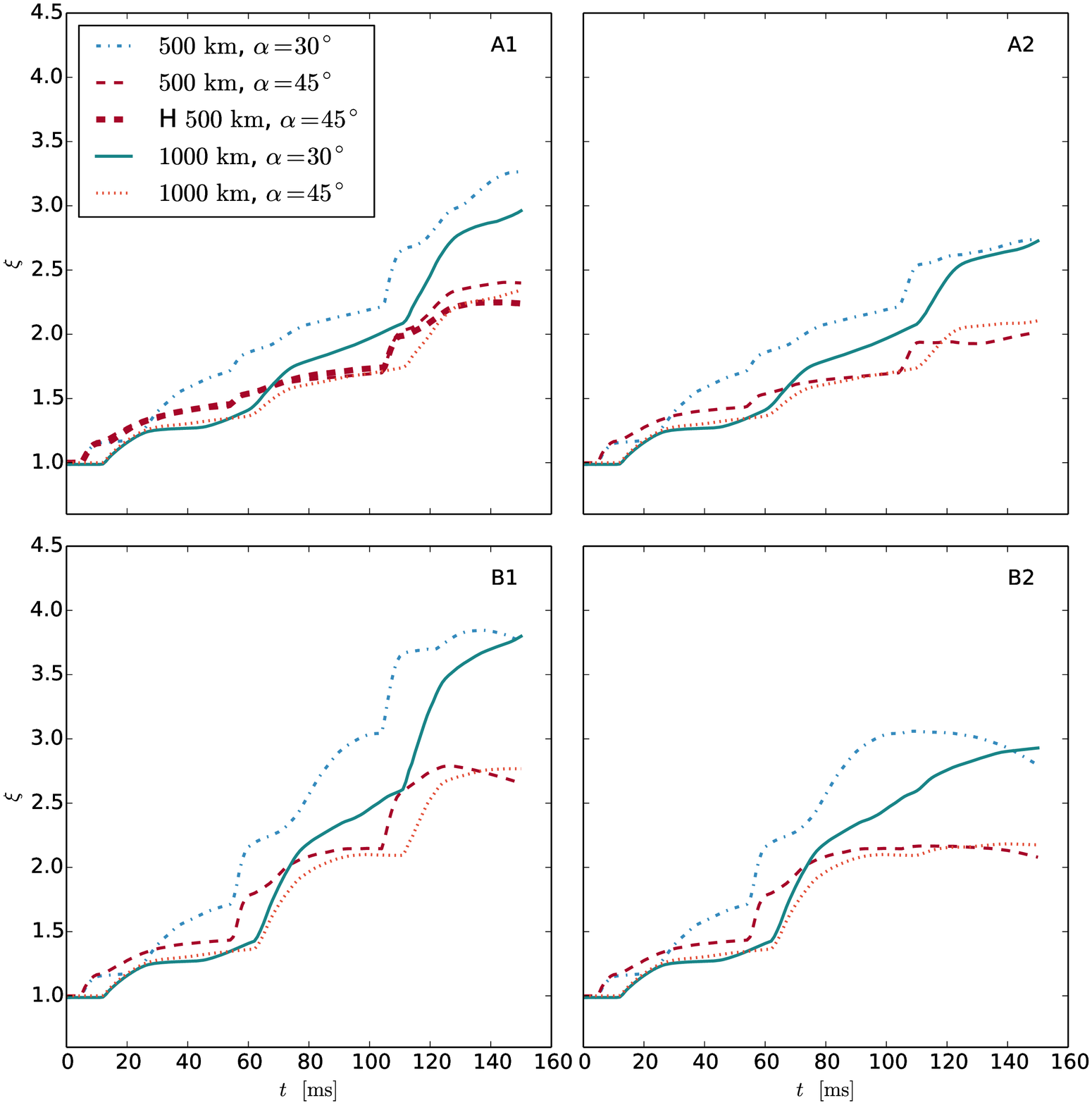}
.
\caption{The normalized mass inflow rate $\xi$, as given in equation \ref{eq:xi1},
within two opposite cones of opening angle $\alpha$ and at two radii.
The four cases marked are according to Table \ref{table}. {{{{The thick line in case A1 is a test run with higher resolution (see section \ref{subsec:res}).} }}} }
\label{fig:acc}
\end{center}
\end{figure}

To further quantify the bipolar accretion pattern we construct two opposite cones whose common vertex is at the center,
and calculate the ratio of the average inflow mass flux through the cones to the average inflow mass flux through the entire sphere at the same radius.
The cones' axis is perpendicular to the $y=0$ plane, i.e., perpendicular to the jets' axes, and through the center.
The quantity we use is
\begin{equation}
\xi \equiv \frac{ \dot M_{\rm cone} ({\rm inflow})/\Omega  }{ M_{\rm total} ({\rm inflow})/ 4 \pi} ,
\label{eq:xi1}
\end{equation}
where $\Omega$ is the solid angle covered by the two cones.
The evolution of $\xi$ with time for two cone-pairs and at two radii are presented in Fig. \ref{fig:acc},
for the 4 different Runs listed in Table \ref{table}. In one cones pair each cone has an opening angle of $30^\circ$, and in the other each cone has an opening angle of $45^\circ$.
We note that the flow becomes more concentrate, i.e., $\xi$ increases with time, to the perpendicular direction
as more jets are launched in the $y=0$ plane. The mass flux per unit solid angle for the $30^\circ$ cones is substantially larger
than that for the $45^\circ$ cones, $\xi(30^\circ) > \xi(45^\circ)$.
This implies that the inflow has the pattern of two opposite stream columns.
This is what we refer to as a bipolar accretion pattern.

\subsection{Resolution dependency of the results}
\label{subsec:res}
\begin{figure}[h]
        \centering
                \centering
                \includegraphics[width=0.8\textwidth]{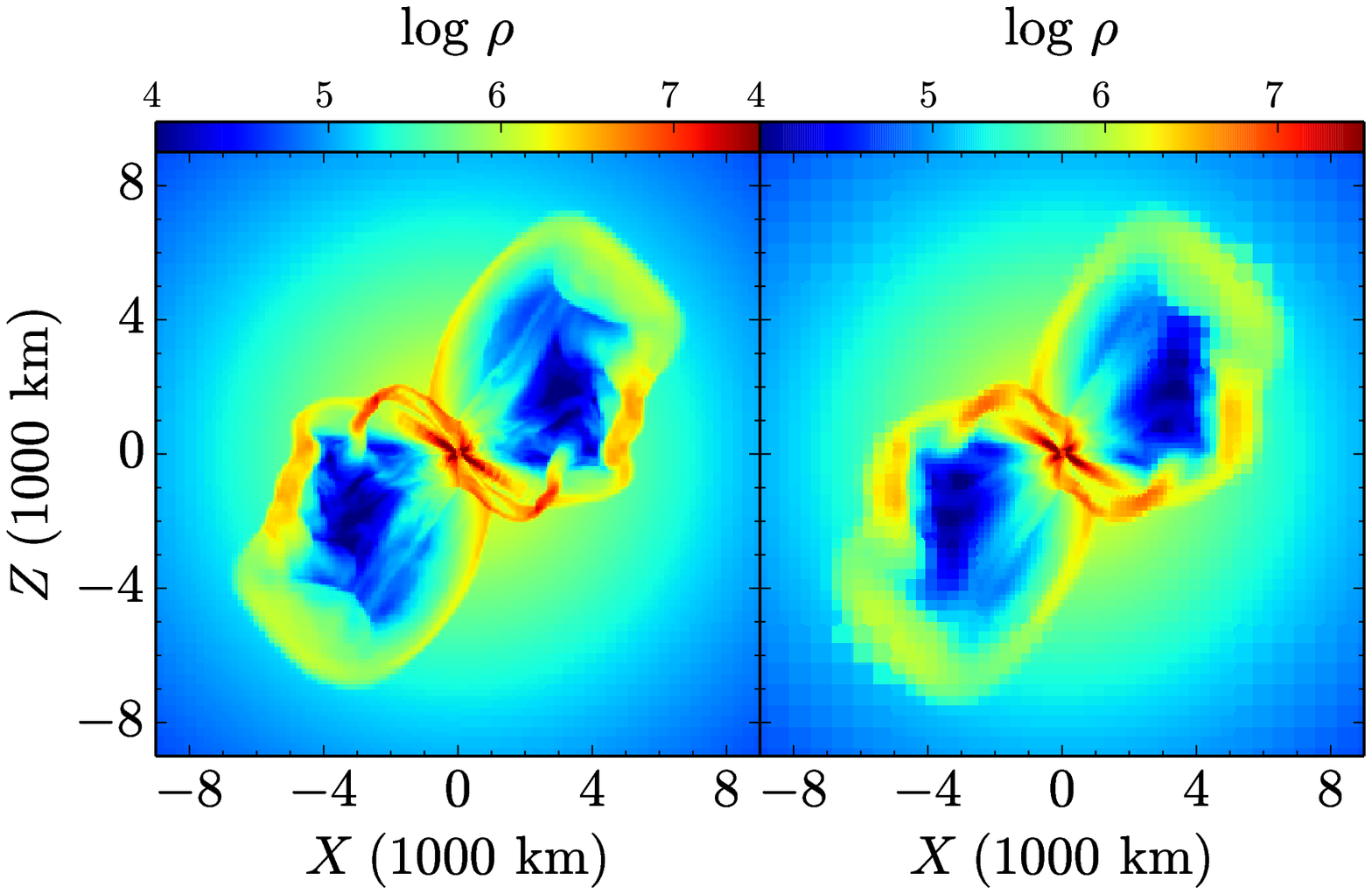}

\caption{{Shown are results for case A1 at time $t=0.15 \s$ with higher resolution (left) and standard resolution (right). Beside the expected sharper features in the high-resolution run, the large scale flow is very similar.}}
\label{fig:high}
\end{figure}

{{{ { To check the dependency of our simulations on resolution we ran a test case of case A1 with a higher resolution (see Fig. \ref{fig:res} for the resolution vs radius). We find the accretion rates in the low and high resolution runs to be within $10 \%$ of each other, as presented by the thick dashed line in the upper-left panel of Fig. \ref{fig:acc}. 
In Fig. \ref{fig:high} we compare the flow structure in the low and high resolution runs at one time.
As expected, the features are sharper in the high-resolution run, but the large-scale flow structure is very similar.   } }}}

We next turn to discuss the implication of the bipolar accretion pattern.

\section{IMPLICATIONS AND SUMMARY}
\label{sec:summary}

In this study we assumed that CCSN explosions are driven by jittering-jets \citep{Papish2011}, an examined the pattern by
which jets are launched.
The angular momentum of the core is not large, such that the jets' axis is determined, at least in part,
by stochastic processes such as instabilities and convection in the pre-collapse core \citep{GilkisSoker2013}.
We conducted 3D numerical simulations with the FLASH code \citep{Fryxell2000}.
In each jets-launching episode we launched one pair of bipolar jets (Fig. \ref{fig:dens-temp-05}).
Under these assumptions, the directions of the two first jets-launching episode are more or less random.
However, the first two episodes, if not along the same direction, define a plane.
We set this plane to be $y=0$ in our study.
The jets will eject mass outward along their propagation direction, leaving inflow in perpendicular directions.
After the first episode the inflow is from a belt region perpendicular to the jets' axis, as
demonstrated in Fig. \ref{fig:one_jets500}.
After the second episode the inflow is concentrated in two opposite directions, as clearly seen in Figs. \ref{fig:NormalPlane}, \ref{fig:second_jets-40} and \ref{fig:second_jets-70}.
A bipolar accretion flow with its axis perpendicular to the $y=0$ plane has been formed.

An inflowing gas along a direction normal to the $y=0$ plane has an angular momentum direction within that plane.
This implies that bipolar accretion flow makes it more likely that the axis of the jets in the third episode will be in
the $y=0$ plane, as jets are launched along the angular momentum axis.
Namely, the third episode jets' axis will be in about the same plane as the axes of the first two episodes.

We simulated two directions for the second jets episode, and for each of these we simulated two directions in the same plane
for a third jets episode. These four cases are summarized in Table \ref{table}.
The asymmetry in the accretion flow is presented for these cases in Fig. \ref{fig:acc}, quantified by the the parameter $\xi$
defined in equation (\ref{eq:xi1}).
The bipolar pattern becomes more prominent as the inflow becomes more concentrated along two opposite directions.
This implies that the following launching of jets will be in, or near, the $y=0$ plane.

The bubbles inflated by the jets grow and expand to all directions as they move toward lower density core gas.
Eventually they will close on directions perpendicular to their axis as well, and expel the outer core and the rest of the star in all directions.
In addition, some of the jets will not be exactly in the same plane, but will have some stochastic variations from the $y=0$ plane.
This will also help in expelling the rest of the star in all directions.
This later evolutionary phase will be studied in the future.

Can this \emph{planar jittering pattern} have any observational consequences?
We don't expect prominent signature in in Type II SNe where the shock will become more spherical as it propagate through
the extended massive hydrogen envelope.
In type Ic SNe the imprint might exist in the supernova remnant (SNR).
We raise here the possibility that the torus morphology of a tilted thick disk with multiple jets in Cassiopeia A SNR (\citealt{DeLaneyetal2010,MilisavljevicFesen2013}) is a result of a planar jittering pattern. 
The last jet was more free to expand, as all core has been removed, and it is now observed as the high velocity jet like outflow along the northeast direction of the SNR.

Finally, we point out that a planar jet launching patter might have taken place during galaxy formation, where a feedback
between accretion of cold gas and jet activity might have taken place.

{{{  {We thank the referee, Jason Nordhaus, for helpful comments.} }}}
This research was supported by the Asher Fund for Space Research at the Technion, and a generous grant from the president of the Technion Prof. Peretz Lavie. OP is supported by the Gutwirth Fellowship. {{{ {The software used in this work was developed in part by the DOE NNSA ASC- and DOE Office of Science ASCR-supported Flash Center for Computational Science at the University of Chicago. This work was supported by the Cy-Tera Project (ΝΕΑ $ \rm Y \Pi O \Delta O M H / \Sigma T P A T H$/0308/31), which is co-funded by the European Regional Development Fund and the Republic of Cyprus through the Research Promotion Foundation.}}}}


\label{lastpage}

\end{document}